\documentstyle[12pt]{article}

\newcommand {\beq}{\begin{equation}}
\newcommand {\eeq}{\end{equation}}
\newcommand {\beqa}{\begin{eqnarray}}
\newcommand {\eeqa}{\end{eqnarray}}         %Equation version
\newcommand {\beqs}{\begin{eqnarray*}}
\newcommand {\eeqs}{\end{eqnarray*}}
\newcommand {\bds}{\begin{displaymath}}
\newcommand {\eds}{\end{displaymath}}
\newcommand {\n}{\nonumber\\}
\newcommand {\nn}{\nonumber}
\newcommand{\no}{\noindent}
\newcommand {\bebb}{\begin{thebibliography}}
\newcommand {\eebb}{\end{thebibliography}}      %Reference version
\newcommand {\bbit}{\bibitem}
\def\al{\alpha}
\def\bt{\beta}

\def\dl{\delta}

\def\gm{\gamma}

\def\Lm{\Lambda}

\def\ph{\phi}

\def\ps{\psi}

\def\tht{\theta}

\def\tl{\tilde}  %%%%%%%%%%For over letters%%%%%%%%%

\def\p{\partial}

\def\psd{\psi ^{\dagger}}

\def\journal#1&#2(#3){\unskip, \sl #1\ \bf #2 \rm(19#3) }
\def\andjournal#1&#2(#3){\sl #1~\bf #2 \rm (19#3) }

\def\npb#1#2#3{Nucl. Phys. {\bf B#1}, (#2) #3}

\def\plb#1#2#3{Phys. Lett. {\bf B#1}, (#2) #3}

\def\cmp#1#2#3{Commun. Math. Phys. {\bf #1}, (#2) #3}

\begin{document}

\begin{titlepage}

\begin{flushright}
\end{flushright}

\vskip 1cm

\begin{center}
%\title
{\Large\bf  $gl(2|2)$ Current Superalgebra and Non-unitary 
   Conformal Field Theory}

\vspace{1cm}

%\author{
{\normalsize\bf
Xiang-Mao Ding $^{a,b}$, %{\thanks{E-mail:xmding@maths.uq.edu.au}}
Mark D. Gould $^a$, Yao-Zhong Zhang $^a$
}
\vskip.1in
{\em $^a$ Centre for Mathematical Physics, Division of Mathematics, \\
University of Queensland, Brisbane, Qld 4072, Australia}
\\
{\em $^b$ Institute of Applied Mathematics, Academy of Mathematics 
and System Sciences; Chinese Academy of Sciences, P.O.Box 2734, 
100080, China.}

\end{center}

\date{}

%\maketitle

\vspace{2cm}

\begin{abstract}
Motivated by application of current superalgebras in the study of
disordered systems such as the random $XY$ and Dirac models, we
investigate $gl(2|2)$ current superalgebra at general level
$k$. We construct its free field representation and corresponding
Sugawara energy-momentum tensor in the non-standard basis. Three screen
currents of the first kind are also presented.

\end{abstract}

\vspace{1cm}

%%%%%PACS: 11.25Hf; 11.30.Rd; 03.65Fd; 02.20.Hj.

\vspace{0.5cm}

\end{titlepage}

\setcounter{section}{0}
\setcounter{equation}{0}
\section{Introduction}

Virasoro algebra and current algebras are algebraic structures in 
conformal field theories (CFTs) \cite{BPZ,Kac,Ph} and 
string theory~\cite{Pol}. According to the Zamolodchikov c-theorem \cite{ZamA}
a unitary CFT corresponds to a 
fixed point of the renormalization group flow. So unitary CFTs 
and current algebras also play an important role in the study of
2D critical behaviours of statistical mechanics models.

Recently current superalgebras and their corresponding non-unitary 
CFTs have attracted much attention. The reason is partly
as follows. One the one hand, representations of superalgebras are often
indecomposable, and such indecomposable representations lead to
logarithmic singularities in correlation functions: a character of
so-called logarithmic CFTs (e.g. see e.g. \cite{Flohr}
and reference therein). On the other hand,
current superalgebras with zero superdimension have found various
physical applications ranging from condensed matter physics to high
energy physics. Particularly interesting are the applications
in topological field theory \cite{IR} and
the supersymmetric method \cite{Efe83} to the study of disordered systems 
\cite{Mud96,BCKT,Ber95,Maa97,Gad91,Gur99}. There the
vanishing of Virasoro central charges and the existence of negative 
dimensional primary fields are essential \cite{Ber95,Mud96,MRF}. 

Free field realization \cite{Wak} is a common approach used in CFTs,
representation theory of current (super)algebras and applications
\cite{BOo,FF,BMP,Ito,Fren,Bow96,Ras98,Din01,DGMZ}. From the viewpoint
of application to disordered systems, most interesting are 
$osp(N|N)$ and $gl(N|N)$ current superalgebras \cite{Ber95,Gur99}. 
The representations of $gl(1|1)$ and $osp(2|2)$ current superalgebras at
general level $k$ have
been studied in details in \cite{RS,Mud96,Maa97,DGMZ}. 

In this paper we investigate the non-semisimple $gl(2|2)$ current
superalgebra at general level $k$, relevant to the two-species random $XY$
and Dirac models. We construct its  free field
representation in the non-standard basis. Let us point out that
a free field realization for the semisimple $sl(2|2)$ 
current superalgebra in the standard basis was obtained in \cite{Ras00}.   
We moreover construct the Sugawara energy-momentum tensor
and obtain three screen currents of the first kind.

\section{Notations}

Unlike bosonic algebras, the simple root system of superalgebras
is not unique. In the case of $gl(N|N)$, the standard (distingished) 
basis has a single fermionic simple root. From physical application
point of view, the non-standard basis where all simple roots are fermionic
is more useful. In this paper, we will adapt the non-standard basis. 
In this basis, $gl(2|2)$ has simple 
raising generators $E_{13}$, $E_{32}$ and $E_{24}$,  and
simple lowering generators $E_{31}$, $E_{23}$ and $E_{42}$. The 
corresponding Cartan subalgebra is generated by
\beqa
H_1=E_{11}+E_{33}, ~~~H_2=E_{22}+E_{33},~~~H_3=E_{22}+E_{44}, \n
H_4=E_{11}+E_{22}-E_{33}-E_{44}+\al(E_{11}+E_{22}+E_{33}+E_{44}),
\eeqa

\no where $\al$ is an arbitrary parameter.  
That $H_4$ is not uniquely determined
is a consequence of the fact that $gl(2|2)$ is
non-semisimple. The defining representation of $gl(2|2)$ in the 
non-standard basis is given by
\beqa
 &&E_{13}= e_{13},~~~ E_{32}= e_{32},~~~E_{24}= e_{24},\n
 &&E_{31}= e_{31}, ~~~E_{23}= e_{23},~~~ E_{42}= e_{42}, \n
 &&E_{12}= e_{12}, ~~~E_{34}=e_{34},~~~E_{14}=e_{14}, \n
 &&E_{21}= e_{21}, ~~~E_{43}=e_{43},~~~E_{41}=e_{41}, \n
 &&H_1=e_{11}+e_{33},~~~H_2=e_{22}+e_{33},~~~H_3=e_{22}+e_{44}, \n
 &&H_4=e_{11}+e_{22}-e_{33}-e_{44}+\al(e_{11}+e_{22}+e_{33}+e_{44}),
\eeqa 

\no where $e_{ij}$ are the matrix with entry $1$ 
at the $i$-th row and $j$-th column, and zero elsewhere,
The generators obey the (anti-) commutation relations:
\beq
 \left[E_{ij}, E_{kl}\right]=\dl _{jk}E_{il}- 
 (-1)^{([i]+[j])([k]+[l])}\dl _{il}E_{jk},
\eeq
where $[1]=[2]=0,~ [3]=[4]=1$ and
\beqa
&&E_{11}=\frac{1}{4}\left( (3-\al)H_1-2H_2+(1-\al)H_3+H_4\right),\n
&&E_{22}=\frac{1}{4}\left(-(1+\al)H_1+2H_2+(1-\al)H_3+H_4\right),\n
&&E_{33}=\frac{1}{4}\left( (1+\al)H_1+2H_2-(1-\al)H_3-H_4\right),\n
&&E_{44}=\frac{1}{4}\left( (1+\al)H_1-2H_2+(3+\al)H_3-H_4\right).
\eeqa

The quadratic Casimir of $gl(2|2)$ is  
\beqa
C_1&=&\sum_{A,B}(-1)^{g(B)}E_{AB}E_{BA} \n
   &=&\frac{1}{2}H_1(H_1-2)-\frac{1}{2}H_3(H_3+2) 
   -(H_1-H_3)H_2+\frac{1}{2}(H_1+H_3)H_4 \n
   & &+2E_{12}E_{21}-2E_{13}E_{31}-2E_{24}E_{42}
   -2E_{14}E_{41}+2E_{32}E_{23}-2E_{34}E_{43} \n
   & &-\frac{\al}{2}(H_1+H_3)^2.
\eeqa
Because $gl(2|2)$ is non-semisimple, $\sum_AE_{AA}$ is a
central element and thus there is another quadratic Casimir
\beq
C_2=\sum_{A,B}E_{AA}E_{BB}
=\left(E_{11}+E_{22}+ E_{33}+ E_{44} \right)^2 \n
=\left(H_1+H_3\right)^2.
\eeq
These two Casimir elements are useful in the following for the
construction of Sugawara energy-momentum tensor.

The $gl(2|2)$ current superalgebra at general level $k$ can be
written as
\beq
J_{ij} (z) J_{kl} (w)=k\,\frac{str(E_{ij}E_{kl})}{(z-w)^2} 
+\frac{1}{z-w}\left( \dl _{jk}J_{il} (w)
-(-1)^{({i}+[j])([k]+[l])}{\dl _{il}}J_{jk} (w)\right),
\label{surpcrt}
\eeq
where $J_{ij}(z)$ are currents corresponding to generators
$E_{ij}$ and
\beqa
J_{11}(z)&=&\frac{1}{4}\left((3-\al)J_1(z)-2J_2(z)
  +(1-\al)J_3(z)+J_4(z)\right), \n
J_{22}(z)&=&\frac{1}{4}\left(-(1+\al)J_1(z)
  +2J_2(z)+(1-\al)J_3(z)+J_4(z)\right), \n
J_{33}(z)&=&\frac{1}{4}\left((1+\al)J_1(z)
  +2J_2(z)-(1-\al)J_3(z)-J_4(z)\right), \n
J_{44}(z)&=&\frac{1}{4}\left((1+\al)J_1(z)-2J_2(z)
  +(3+\al)J_3(z)-J_4(z)\right). \nn
\eeqa
Here $J_i(z)$ are currents associated with $H_i$.

\section{Free field realization}

To obtain a free field realization, 
we first construct Fock space representation of $gl(2|2)$. 
The Fock space is constructed by the 
actions of the raising operators
$E_{13}$, $E_{32}$, $E_{24}$, $E_{12}$, $E_{34}$ 
and $E_{14}$ on the lowest weight state. Define the lowest 
weight state $|\Lm>$ of $gl(2|2)$ by  
\beqa
&&E_{13}|\Lm >=E_{32}|\Lm >=E_{24}|\Lm >
=E_{12}|\Lm >=E_{34}|\Lm >=E_{14}|\Lm >=0, \n
&&H_i |\Lm >={\Lm }_i |\Lm >.
\eeqa
\no Then in the non-standard basis, the action of the 
operator $e^{A}$ with 
\[
A={\tht}_{13}E_{13}+{\tht}_{32}E_{32}+{\tht}_{24}E_{24}
+{x}_{12}E_{12}+{x}_{34}E_{34}+{\tht}_{14}E_{14}
\]
on the lowest weight state $|\Lm >$ generates a coherent state 
of the algebra, where $x_{ij}$ are bosonic coordinates
satisfying $x_{12}x_{34}=x_{34}x_{12}$ and 
$[\p _{x_{ij}}, x_{kl}]=\dl _{ik}\dl _{jl}$ and $\tht _{ij}$ 
are fermionic coordinates obeying
$\tht _{ij}\tht _{kl}=-\tht _{kl}\tht _{ij}$ and 
$\{\p _{\tht _{ij}}, \tht _{kl}\}=\dl _{ik}\dl _{jl}$. 

We write
\beq
ge^{A}|\Lm>=d_ge^{A}|\Lm>,
\eeq 
\no where $g$ s a generator of $gl(2|2)$ and 
$d_g$ is the corresponding differential operator. 
By using the defining relations of $gl(2|2)$ and the 
Baker-Campbell-Hausdorff formula, after a long algebraic
manipulation we obtain
\beqa
&&d_{E_{14}}=\p _{\tht_{14}}, \n
&&d_{E_{12}}=\p _{x_{12}}-\frac{1}{2}{\tht}_{24}\p _{\tht_{14}}, \n
&&d_{E_{34}}=\p _{x_{34}}+\frac{1}{2}{\tht}_{13}\p _{\tht_{14}}, \n
&&d_{E_{13}}=\p _{\tht_{13}}+\frac{1}{2}\tht_{32}\p _{x_{12}}
            -\frac{1}{2}x_{34}\p _{\tht_{14}}
            -\frac{1}{12}{\tht}_{32}{\tht}_{24}\p _{\tht_{14}}, \n
&&d_{E_{32}}=\p _{\tht_{32}}
            +\frac{1}{2}\tht_{13}\p _{x_{12}}
	    +\frac{1}{2}\tht_{24}\p _{x_{34}}
	    -\frac{1}{6}{\tht}_{13}{\tht}_{24}\p _{\tht_{14}}, \n
&&d_{E_{24}}=\p _{\tht_{24}}
	    +\frac{1}{2}\tht_{32}\p _{x_{34}}
	    +\frac{1}{2}x_{12}\p _{\tht_{14}}
	    +\frac{1}{12}{\tht}_{32}{\tht}_{13}\p _{\tht_{14}}, \n
&&d_{H_1}=\Lm _1
         -\tht _{32}\p _{\tht_{32}}
 	 -x _{12}\p _{x_{12}}
	 -x _{34}\p _{x_{34}}
	 -\tht _{14}\p _{\tht_{14}}, \n
&&d_{H_2}=\Lm_2
         +\tht _{13}\p _{\tht_{13}}
 	 -\tht _{24}\p _{\tht_{24}}
	 +x _{12}\p _{x_{12}}
	 -x _{34}\p _{x_{34}}, \n
&&d_{H_3}=\Lm_3
          +\tht _{32}\p _{\tht_{32}}
          +x _{12}\p _{x_{12}}
	  +x _{34}\p _{x_{34}}
	  +\tht _{14}\p _{\tht_{14}},\n 
&&d_{H_4}=\Lm_4
          -2\left(\tht _{13}\p _{\tht_{13}}
                 -\tht _{32}\p _{\tht_{32}}
                 +\tht _{24}\p _{\tht_{24}}
		 +\tht _{14}\p _{\tht_{14}}\right),\n 
&&d_{E_{31}}={\tht_{13}}\Lm_1 
   -x_{12}\p _{\tht_{32}}
   +\tht _{14} \p _{x_{34}} 
   -\frac{1}{2}\tht_{13}
   \left(
   \tht_{32}\p _{\tht_{32}}
   + x_{12}\p _{x_{12}}
   +x_{34}\p _{x_{34}}
   +\tht_{14}\p _{\tht_{14}} \right) \n
   &&~~~~~~~~+\frac{1}{12} \tht_{13} 
   \left(  
   {\tht}_{32}{\tht}_{24}\p _{x_{34}}
   -x_{12}{\tht}_{24}\p _{x_{34}}
   \right), \n
&&d_{E_{23}}={\tht_{32}}\Lm_2 
   +x_{12}\p _{\tht_{13}}
   -x _{34} \p _{\tht_{24}} 
   +\frac{1}{2}\tht_{32}
   \left(
   \tht_{13}\p _{\tht_{13}}
   - \tht_{24}\p _{\tht_{24}}
   + x_{12}\p _{x_{12}}
   -x_{34}\p _{x_{34}} \right) \n
   &&~~~~~~~~+\frac{1}{6} \tht_{32} 
   \left(  
   {\tht}_{13}x_{34}\p _{\tht_{14}}
   + x_{12}{\tht}_{24}\p _{\tht_{14}}
   \right), \n
&&d_{E_{42}}={\tht_{24}}\Lm_3 
   +x_{34}\p _{\tht_{32}}
   +\tht _{14} \p _{x_{12}} 
   +\frac{1}{2}\tht_{24}
   \left(
   \tht_{32}\p _{\tht_{32}}
   + x_{12}\p _{x_{12}}
   +x_{34}\p _{x_{34}} 
   + \tht_{14}\p _{\tht_{14}} \right) \n
   &&~~~~~~~~+\frac{1}{12} \tht_{24} 
   \left(  
   {\tht}_{13} \tht_{32}\p _{x_{12}}
   - {\tht}_{13}x_{34}\p _{\tht_{14}}
   \right), \n
&&d_{E_{21}}=(x_{12} -\frac{1}{2}{\tht}_{13}\tht_{32})\Lm_1
            -(x_{12} +\frac{1}{2}{\tht}_{13}\tht_{32})\Lm_2 
   -x_{12} \left(
   {\tht}_{13}\p _{\tht_{13}} 
   +\tht_{32}\p _{\tht_{32}}
   + x_{12}\p _{x_{12}} \right) \n
   &&~~~~~~~~-\left(
   \tht _{14}-\frac{1}{2}x_{12}\tht_{24}
   -\frac{1}{2}\tht_{13}x_{34} 
   -\frac{1}{6}\tht_{13}\tht_{32}\tht_{24} 
   \right) \p _{\tht _{24}} \n
   &&~~~~~~~~+\frac{1}{2}\tht_{32}
   \left(
   \tht _{14}+\frac{1}{2}x_{12}\tht_{24}
   +\frac{1}{2}\tht_{13}x_{34} 
   \right)\p _{x _{34}} \n
   &&~~~~~~~~-\frac{1}{2} x_{12}
   \left(
   \tht _{14}+\frac{1}{2}x_{12}\tht_{24}
   +\frac{1}{2}\tht_{13}x_{34} 
   \right)\p _{\tht _{14}}
   +\frac{1}{12}\tht_{13}\tht_{32}
   \left(
   \tht _{14}-\frac{1}{2}x_{12}\tht_{24}
   \right)\p _{\tht _{14}}, \n
&&d_{E_{43}}=(x_{34} -\frac{1}{2}{\tht}_{32}\tht_{24})\Lm_2
            -(x_{34} +\frac{1}{2}{\tht}_{32}\tht_{24})\Lm_3 
   -x_{34} \left(
   \tht_{32}\p _{\tht_{32}}
   +{\tht}_{24}\p _{\tht_{24}}
   +x_{34}\p _{x_{34}} \right) \n
   &&~~~~~~~~-\left(
   \tht _{14}+\frac{1}{2}x_{12}\tht_{24}
   +\frac{1}{2}\tht_{13}x_{34} 
   -\frac{1}{6}\tht_{13}\tht_{32}\tht_{24} 
   \right) \p _{\tht _{13}} \n
   &&~~~~~~~~-\frac{1}{2}\tht_{32}
   \left(
   \tht _{14}+\frac{1}{2}x_{12}\tht_{24}
   -\frac{1}{2}\tht_{13}x_{34} 
   \right)\p _{x _{34}} \n
   &&~~~~~~~~-\frac{1}{2} x_{34}
   \left(
   \tht _{14}-\frac{1}{2}x_{12}\tht_{24}
   -\frac{1}{2}\tht_{13}x_{34} 
   \right)\p _{\tht _{14}}
   -\frac{1}{12}\tht_{32}\tht_{24}
   \left(
   \tht _{14}+\frac{1}{2}\tht_{13}x_{34}
   \right)\p _{\tht _{14}}, \n
&&d_{E_{41}}=\left(
   \tht _{14}+\frac{1}{2}x_{12}\tht_{24}
   +\frac{1}{2}\tht_{13}x_{34} 
   -\frac{1}{6}\tht_{13}\tht_{32}\tht_{24}
   \right)\Lm_1 \n
   &&~~~~~~~~-\left(
   \tht _{14}+\frac{1}{2}x_{12}\tht_{24}
   -\frac{1}{2}\tht_{13}x_{34} 
   +\frac{1}{3}\tht_{13}\tht_{32}\tht_{24} 
   \right)\Lm_2 \n
   &&~~~~~~~~+\left(
   \tht _{14}-\frac{1}{2}x_{12}\tht_{24}
   -\frac{1}{2}\tht_{13}x_{34} 
   -\frac{1}{6}\tht_{13}\tht_{32}\tht_{24} 
   \right)\Lm_3 \n
   &&~~~~~~~~+\tht _{13} \left(
   \tht _{14}+\frac{1}{2}x_{12}\tht_{24}
   \right)\p _{\tht _{13}}
    -\tht _{24}\left(
   \tht _{14}-\frac{1}{2}\tht_{13}x_{34} 
   \right)\p _{\tht _{24}} \n
   &&~~~~~~~~-\left(
   x_{12}x_{34} 
   -\frac{1}{2}x_{12}\tht_{32}\tht_{24}
   +\frac{1}{2}\tht_{13}\tht_{32}x_{34} 
   \right)\p _{\tht _{32}} \n
   &&~~~~~~~~-x_{12}
   \left(
   \tht _{14}+\frac{1}{2}x_{12}\tht_{24}
   +\frac{1}{12}\tht_{13}\tht_{32}\tht_{24}
   \right)\p _{x_{12}} \n
   &&~~~~~~~~+x_{34}   
   \left(
   \tht _{14}-\frac{1}{2}\tht_{13}x_{34} 
   +\frac{1}{12}\tht_{13}\tht_{32}\tht_{24}
   \right)\p _{x_{34}}
   +\frac{1}{6}x_{12}\tht_{24}\tht_{13}x_{34}
   \p _{\tht _{14}}.
\eeqa  

\no It is straightforward to prove that the above differential 
operators satisfy the algebraic relations of $gl(2|2)$ algebra. 
From the differential operators realization, we see that the 
parameter $\al$ involved term has no contribution to the result.  

With the help of the differential operator representation 
we can find the Wakimoto realization of $gl(2|2)$ current
superalgebra in terms of two bosonic $\bt$-$\gm$ pairs, four
fermionic $b$-$c$ type systems and four free scalar fields $\ph_i$.
The free fields obey the following OPEs:
\beqa
&&\bt _{ij}(z)\gm _{kl}(w)=-\gm_{kl}(z)\bt_{ij}(w)=
   \frac{\dl _{ik}\dl _{jl}}{z-w},\n
&&\ps _{ij}(z)\psd _{kl}(w)=\psd_{kl}(z)\ps_{ij}(w)=
   \frac{\dl _{ik}\dl _{jl}}{z-w},\n
&&\ph _i (z)\ph _j (w)=-str(H_iH_j)\,\ln(z-w).
\eeqa
The free field realization of the $gl(2|2)$ current
superalgebra is obtained by the substitution:
\beqa
&&d_{E_{ij}}\rightarrow J_{ij}(z),~~~~d_{H_i}\rightarrow
J_i(z),~~~~
x_{ij}\rightarrow \gm_{ij}(z),~~~~\partial_{x_{ij}}\rightarrow
\beta_{ij}(z),\n
&&\theta_{ij}\rightarrow \ps_{ij}(z),~~~~
\partial_{\theta_{ij}}\rightarrow \psd_{ij}(z),~~~~
\Lambda_i\rightarrow \sqrt{-k}\partial\phi_i(z)
\eeqa
in the differential operator realization of $gl(2|2)$ and a
subsequent addition of anomalous terms linear in $\partial
\psi_{ij}(z)$ and $\partial\gamma_{ij}(z)$ in currents
associated with the lowering generators of $gl(2|2)$. It also
turns out that one needs to add the ``anomalous" term
$\p \ph _1(z) + \p \ph _3(z)$ with an appropriate coefficient
to the current $J_4(z)$. The result is 
\beqa
&&J_{14}(z)=\psd _{14} (z), \n
&&J_{12}(z)=\bt _{12} (z)-\frac{1}{2}\ps _{24} (z)\psd _{14} (z), \n
&&J_{34}(z)=\bt _{34} (z)+\frac{1}{2}\ps _{13} (z)\psd _{14} (z), \n
&&J_{13}(z)=\psd _{13} (z)+\frac{1}{2}\ps _{32} (z)\bt _{12}(z)
           -\frac{1}{2}\left( \gm _{34} (z)
           +\frac{1}{6}{\ps}_{32}(z){\ps}_{24}(z)\right)
	   \psd _{14}(z), \n
&&J_{32}(z)=\psd _{32} (z)
                +\frac{1}{2}\ps _{13}(z)\bt _{12}(z)
                +\frac{1}{2}\ps _{24}(z)\left(\bt _{34}(z)
                +\frac{1}{3}{\ps}_{13} (z)\psd _{14}(z)\right), \n
&&J_{24}(z)=\ps _{24}(z)
                +\frac{1}{2}\ps _{32}(z)\bt _{34}(z)
                +\frac{1}{2}\left(\gm _{12}(z)
                -\frac{1}{6}{\ps}_{13}(z){\ps}_{32}(z)\right)
		\psd _{14}(z), \n
&&J_{1}(z)=i\sqrt{k}\p \ph _1(z)
             -\ps _{32}(z)\psd _{32}(z)
             -\bt _{12}(z)\gm _{12} (z)
	     -\bt _{34}(z)\gm _{34} (z)
	     -\ps _{14}(z)\psd _{14}(z), \n
&&J_{2}(z)=i\sqrt{k}\p \ph _2(z)
             +\ps _{13}(z)\psd _{13}(z)
             -\ps _{24}(z)\psd _{24}(z)
	     +\bt _{12}(z)\gm _{12}(z)
	     -\bt _{34}(z)\gm _{34}(z), \n
&&J_{3}(z)=i\sqrt{k}\p \ph _3(z)
             +\ps _{32}(z)\psd _{32}(z)
             +\bt _{12}(z)\gm _{12}(z)
	     +\bt _{34}(z)\gm _{34}(z)
	     +\ps _{14}(z)\gm _{14}(z),\n 
&&J_{4}(z)=i\sqrt{k}\p \ph _4(z)
          -\frac{2i}{\sqrt{k}}
	  \left(\p \ph _1(z) + \p \ph _3(z)\right) \n
          &&~~~~~~~~-2\left(
	          \ps _{13}(z)\psd _{13}(z)
                 -\ps _{32}(z)\psd _{32}(z)
                 +\ps _{24}(z)\psd _{24}(z)
		 +\ps _{14}(z)\psd _{14}(z)\right),\n
&&J_{31}(z)=i\sqrt{k}\p \ph _1(z){\ps_{13}}(z) 
   -\gm _{12}(z)\psd _{32}(z)
   +\ps _{14}(z) \bt _{34}(z) \n 
   &&~~~~~~~~-\frac{1}{2}\ps _{13}(z) 
   \left(
   \ps _{32}(z)\psd _{32}(z)
   + \bt _{12}(z)\gm _{12}(z)
   +\bt _{34}(z)\gm _{34}(z)
   +\ps _{14}(z)\psd _{14}(z) \right) \n
   &&~~~~~~~~+\frac{1}{12} \ps _{13}(z) 
   \left(  
   {\ps}_{32}(z){\ps}_{24}(z)\bt _{34}
   -\gm _{12}(z){\ps}_{24}(z)\bt _{34}
   \right)
   +k\p {\ps_{13}}(z), \n
&&J_{23}(z)=i\sqrt{k}\p \ph _2(z){\ps_{32}}(z) 
   +\gm _{12}(z)\psd _{13}(z)
   -\gm _{34}(z)\psd _{24}(z) \n 
   &&~~~~~~~~+\frac{1}{2}\ps_{32}(z)
   \left(
   \ps_{13}(z)\psd _{13}(z)
   - \ps_{24}(z)\psd _{24}(z)
   +\bt _{12}(z)\gm _{12}(z)
   -\bt _{34}(z)\gm _{34}(z) \right) \n
   &&~~~~~~~~+\frac{1}{6} \ps_{32}(z) 
   \left(  
   {\ps}_{13}(z)\gm _{34}(z)\psd _{14}(z)
   + \gm _{12}(z){\ps}_{24}(z)\psd _{14}(z)
   \right)
   -k \p {\ps_{32}}(z), \n
&&J_{42}(z)=i\sqrt{k}\p \ph _3(z){\tht_{24}} 
   +\gm _{34}(z)\psd _{32}(z)
   +\ps _{14}(z) \bt _{12}(z) \n
   &&~~~~~~~~+\frac{1}{2}\ps_{24}(z)
   \left(
   \ps_{32}(z)\psd _{32}(z)
   +\bt_{12}(z)\gm _{12}(z)
   +\bt_{34}(z)\gm _{34}(z) 
   +\ps_{14}(z)\psd _{14}(z) \right) \n
   &&~~~~~~~~+\frac{1}{12} \ps_{24}(z) 
   \left(  
   {\ps}_{13}(z) \ps_{32}(z)\bt _{12}(z)
   -{\ps}_{13}(z)\gm _{34}(z)\psd _{14}(z)
   \right) 
   +k\p {\ps_{24}}(z), \n
&&J_{21}(z)=i\sqrt{k}\p \ph _1(z)
               (\gm_{12}(z) -\frac{1}{2}{\ps}_{13}(z)\ps_{32}(z))
               -i\sqrt{k}\p \ph _2(z)
	       (\gm _{12}(z) +\frac{1}{2}{\ps}_{13}(z)\ps_{32}(z)) \n
   &&~~~~~~~~-\gm _{12}(z) \left(
   {\ps}_{13}(z)\psd _{13}(z) 
   +\ps_{32}(z)\psd _{32}(z)
   +\bt_{12}(z)\gm _{12}(z) \right) \n
&&~~~~~~~~-\left(
   \ps _{14}(z)-\frac{1}{2}\gm_{12}(z)\ps_{24}(z)
   -\frac{1}{2}\ps_{13}(z)\gm _{34}(z) 
   -\frac{1}{6}\ps _{13}(z)\ps_{32}(z)\ps_{24}(z) 
   \right) \psd _{24}(z) \n
&&~~~~~~~~+\frac{1}{2}\ps_{32}(z)
   \left(
   \ps _{14}(z)+\frac{1}{2}\gm _{12}(z)\ps _{24}(z)
   +\frac{1}{2}\ps _{13}(z)\gm _{34}(z) 
   \right)\bt _{34}(z) \n
&&~~~~~~~~-\frac{1}{2} \gm _{12}(z)
   \left(
   \ps _{14}(z)+\frac{1}{2}\gm_{12}(z)\ps_{24}(z)
   +\frac{1}{2}\ps_{13}(z)\gm_{34}(z) 
   \right)\psd _{14}(z) \n
   &&~~~~~~~~
   +\frac{1}{12}\ps_{13}(z)\ps_{32}(z)
   \left(
   \ps _{14}(z)-\frac{1}{2}\gm_{12}(z)\ps_{24}(z)
   \right)\psd _{14}(z) +(k-1/2)\p \gm _{12}(z)\n
   &&~~~~~~~~+\frac{1}{2}(k+1/6)\ps_{32}(z) \p \ps_{13}(z)
   +\frac{1}{2}(k+1/3)\ps_{13}(z) \p \ps_{32}(z), \n
&&J_{43}(z)=i\sqrt{k}\p \ph _2(z)
             (\gm_{34}(z) -\frac{1}{2}{\ps}_{32}(z)\ps_{24}(z))
            -i\sqrt{k}\p \ph _3(z)
	    (\gm_{34}(z) +\frac{1}{2}{\ps}_{32}(z)\ps_{24}(z))\n
   &&~~~~~~~~-\gm_{34}(z) \left(
             \ps_{32}(z)\psd_{32}(z)
	     +{\ps}_{24}(z)\psd _{24}(z)
             +\bt_{34}(z)\gm _{34}(z) \right) \n
&&~~~~~~~~-\left(
   \ps _{14}(z)+\frac{1}{2}\gm_{12}(z)\ps_{24}(z)
   +\frac{1}{2}\ps_{13}(z)\gm_{34}(z) 
   -\frac{1}{6}\ps_{13}(z)\ps_{32}(z)\ps_{24}(z) 
   \right) \psd _{13}(z) \n
&&~~~~~~~~-\frac{1}{2}\ps_{32}(z)
   \left(
   \ps _{14}(z)+\frac{1}{2}\gm_{12}(z)\ps_{24}(z)
   -\frac{1}{2}\ps_{13}(z)\gm_{34}(z) 
   \right)\bt _{34}(z) \n
&&~~~~~~~~-\frac{1}{2} \gm_{34}(z)
   \left(
   \ps _{14}(z)-\frac{1}{2}\gm_{12}(z)\ps_{24}(z)
   -\frac{1}{2}\ps_{13}(z)\gm_{34}(z) 
   \right)\psd _{14}(z) \n
   &&~~~~~~~~-\frac{1}{12}\ps_{32}(z)\ps_{24}(z)
   \left(
   \ps _{14}(z)+\frac{1}{2}\ps_{13}(z)\gm_{34}(z)
   \right)\psd _{14}(z)
   -(k+1/2)\p \gm_{34}(z) \n
   &&~~~~~~~~-\frac{1}{2}(k-1/3)\ps_{24}(z) \p \ps_{32}(z)
             -\frac{1}{2}(k-1/6)\ps_{32}(z) \p \ps_{24}(z),\n
&&J_{41}(z)=i\sqrt{k}\p \ph _1(z)\left(
   \ps _{14}(z)+\frac{1}{2}\gm_{12}(z)\ps_{24}(z)
   +\frac{1}{2}\ps_{13}(z)\gm_{34}(z) 
   -\frac{1}{6}\ps_{13}(z)\ps_{32}(z)\ps_{24}(z)
   \right) \n
   &&~~~~~~~~-i\sqrt{k}\p \ph _2(z)\left(
   \ps _{14}(z)+\frac{1}{2}\gm_{12}(z)\ps_{24}(z)
   -\frac{1}{2}\ps_{13}(z)\gm_{34}(z) 
   +\frac{1}{3}\ps_{13}(z)\ps_{32}(z)\ps_{24}(z) 
   \right) \n
   &&~~~~~~~~+i\sqrt{k}\p \ph _3(z)\left(
   \ps _{14}(z)-\frac{1}{2}\gm_{12}(z)\ps_{24}(z)
   -\frac{1}{2}\ps_{13}(z)\gm_{34}(z) 
   -\frac{1}{6}\ps_{13}(z)\ps_{32}(z)\ps_{24}(z) 
   \right) \n
   &&~~~~~~~~+\ps _{13}(z) \left(
   \ps _{14}(z)+\frac{1}{2}\gm_{12}(z)\ps_{24}(z)
   \right)\psd _{13}(z)
   +\frac{1}{6}\gm_{12}(z)\ps_{24}(z)\ps_{13}(z)\gm_{34}(z)
   \psd_{14}(z)\n
   &&~~~~~~~~-\left(
   \gm_{12}(z)\gm_{34}(z) 
   -\frac{1}{2}\gm_{12}(z)\ps_{32}(z)\ps_{24}(z)
   +\frac{1}{2}\ps_{13}(z)\ps_{32}(z)\gm_{34}(z) 
   \right)\psd_{32}(z) \n
   &&~~~~~~~~-\gm_{12}(z)
   \left(
   \ps _{14}(z)+\frac{1}{2}\gm_{12}(z)\ps_{24}(z)
   +\frac{1}{12}\ps_{13}(z)\ps_{32}(z)\ps_{24}(z)
   \right)\bt _{12} \n
   &&~~~~~~~~+\gm_{34}(z)   
   \left(
   \ps _{14}(z)-\frac{1}{2}\ps_{13}(z)\gm_{34}(z) 
   +\frac{1}{12}\ps_{13}(z)\ps_{32}(z)\ps_{24}(z)
   \right)\bt_{34}(z) \n
   &&~~~~~~~~-\ps _{24}(z)\left(
   \ps _{14}(z)-\frac{1}{2}\ps_{13}(z)\gm_{34}(z) 
   \right)\psd_{24}(z)
   +\frac{1}{2}(k-1)\ps_{24}(z)\p \gm_{12}(z) \n 
   &&~~~~~~~~-\frac{k}{2}\gm_{12}(z)\p \ps_{24}(z)
   -\frac{1}{2}(k+1)\ps_{13}(z)\p \gm_{34}(z)
   +\frac{k}{2}\gm_{34}(z)\p \ps_{13}(z) \n
   &&~~~~~~~~-\frac{1}{6}(k+1/2)
    \ps_{32}(z)\ps_{24}(z) \p \ps_{13}(z) 
   -\frac{k}{3}\ps_{13}(z)\ps_{24}(z) \p \ps_{32}(z) \n
   &&~~~~~~~~-\frac{1}{6}(k-1/2)\ps_{13}(z)
   \ps_{32}(z) \p \ps_{24}(z)
   +k \p \ps _{14}(z).
\eeqa  

We remark that 
even though the parameter $\al$ does not occur in the expression 
of the currents, it appears in the following OPE 
\[
J_{4}(z)J_{4}(w)=k\frac{8\al}{(z-w)^2}+\ldots.
\]
So different choice of $\al$ will give different OPE of $J_4(z)$ with
itself. It is straightforward to check that the 
above currents satisfy the OPEs of the $gl(2|2)$ current
superalgebra given in the last section.

\section{Energy-momentum tensor} 

The CFT associated with the $gl(2|2)$ current superalgebra is
obtained by constructing the Sugawara energy-momentum tensor.
The Sugawara tensor corresponding to the quadratic Casimir
$C_1$ is given by 
\beqa
&&T_1(z)=\frac{1}{2k}:\sum_{i,j}(-1)^{g(j)}J_{ij}(z)J_{ji}(z):\n
        &&~~~~~~~~
	=\bt _{12}(z) \p \gm _{12}(z) 
        +\bt _{34}(z) \p \gm _{34}(z)
	-\psd _{13}(z) \p \ps _{13}(z)  \n
	&&~~~~~~~~
	-\psd _{32}(z)\p \ps _{32}(z)
        -\psd _{24}(z)\p \ps _{24}(z)
        -\psd _{14}(z)\p \ps _{14}(z)\n
        &&~~~~~~~~
	-\frac{1}{4}\left ( 
	(\p \ph _1(z))^2 -(\p \ph _3(z))^2 
	+\left(\p \ph _1(z)+\p \ph _3(z)\right)\p \ph _4(z)
	\right) \n
        &&~~~~~~~~
	+\frac{1}{2}\left ( 
	\p \ph _1(z) -\p \ph _3(z)
	\right)\p \ph _2 (z)
        +\frac{i}{2\sqrt{k}}\left ( 
	\p ^2\ph _1(z) +\p ^2\ph _3(z)
	\right) \n
	&&~~~~~~~~
	+\frac{4+k\al}{4k}\left ( 
	\p \ph _1(z) +\p \ph _3(z) \right)^2.
\eeqa
With respect to this tensor, all currents except $J_4(z)$  are
primary fields. However, its OPE with $J_4(z)$ reads
\beqa
T_1(z)J_{4}(w)&=&\frac{1}{(z-w)^2}J_{4}(w)
                  +\frac{1}{(z-w)}\p J_{4} \n
		  & &-\frac{1}{(z-w)^2}\frac{4i}{\sqrt k}
		  \left(\p \ph _1(w) +\p \ph _3(w)\right) \n
		  & &-\frac{1}{(z-w)} \frac{4i}{\sqrt k}
		  \left(\p ^2\ph _1(w) +\p ^2\ph _3(w)\right).
\eeqa 
This means that $T_1(z)$ is not a corrent energy-momentum tensor
of the theory.

Now the Sugawara tensor associated with the Casimir $C_2$ is
\beq
T_2(z)=\frac{1}{2k}:\sum_{i,j}J_{ii}(z)J_{jj}(z):
=-\frac{1}{2}\left ( \p \ph _1(z) +\p \ph _3(z)\right)^2.
\eeq
It has a non-trivial OPE with current $J_4(z)$:
\beq
T_2(z)J_4(w)=\frac{4i{\sqrt k}}{(z-w)^2}
		  \left(\p \ph _1(w) +\p \ph _3(w)\right)
		 +\frac{4i{\sqrt k}}{(z-w)}
		  \left(\p ^2\ph _1(w) +\p ^2\ph _3(w)\right).
\eeq
Comparing the two above OPEs, we see that if we define 
\beq
T(z)=T_1(z)+\frac{1}{k}T_2(z)
\eeq
then all currents become primary with respect to $T(z)$ and
moreover
\beq
T(z)T(w)=\frac{c/2}{(z-w)^4}+\frac{2T(w)}{(z-w)^2}
+\frac{\p T(w)}{z-w)},
\eeq
where the central charge $c=0$. So $T(z)$ is the
energy-momentum tensor of the $gl(2|2)$ current superalgebra.
In terms of the free fields $T(z)$ reads
\beqa
&&T(z)=\bt _{12}(z) \p \gm _{12}(z) 
        +\bt _{34}(z) \p \gm _{34}(z)
	-\psd _{13}(z) \p \ps _{13}(z)  \n
	&&~~~~~~~~
	-\psd _{32}(z)\p \ps _{32}(z)
        -\psd _{24}(z)\p \ps _{24}(z)
        -\psd _{14}(z)\p \ps _{14}(z)\n
        &&~~~~~~~~
	-\frac{1}{4}\left ( 
	(\p \ph _1(z))^2 -(\p \ph _3(z))^2 
	+(\p \ph _1(z)+\p \ph _3(z))\p \ph _4(z)
	\right) \n
        &&~~~~~~~~
	+\frac{1}{2}\left ( 
	\p \ph _1(z) -\p \ph _3(z)
	\right)\p \ph _2 (z)
        +\frac{i}{2\sqrt{k}}\left ( 
	\p ^2\ph _1(z) +\p ^2\ph _3(z)
	\right)\n
	&&~~~~~~~~
	+\frac{\al}{4}\left ( 
	\p \ph _1(z) +\p \ph _3(z) \right)^2.
\eeqa

\section{Screening currents}
 
An important object in the free field approach is screening current. 
Screening currents are primary fields with conformal dimension $1$, 
and their integrations give the screening charges. Screening
currents  communite with icurrent generators
up to a total derivative. These properties ensure that 
screening charges may be inserted into correlators while the 
conformal or affine ward identities remain intact. 
For the present case, we find three screening currents,
\beqa
s_{13}(z)&=&\left(
  -\psd _{13}(z)+\frac{1}{2}\bt _{12}(z) \ps _{32}(z)
  -\frac{1}{2}\left(
  \gm _{34}(z)-\frac{1}{6}\ps _{32}(z)\ps _{24}(z)
  \right)\psd _{14}(z)
  \right)
{\tl s_{13}}(z), \n
s_{32}(z)&=&\left(
   -\psd _{32}(z)+\frac{1}{2}\bt _{12}(z) \ps _{13}(z)
   +\frac{1}{2}\ps _{24}(z)
   \left(\bt _{34}(z)-\frac{1}{3}\ps _{13}(z)\psd _{14}(z)
   \right)
   \right)
   {\tl s_{32}}(z), \n
s_{24}(z)&=&\left(
   -\psd _{24}(z)+\frac{1}{2}\bt _{34}(z) \ps _{32}(z)
   +\frac{1}{2}
   \left(
   \gm _{12}(z)+\frac{1}{6}\ps _{13}(z)\ps _{32}(z)
   \right)\psd _{14}(z)
   \right)
   {\tl s_{24}}(z),\n 
\eeqa
where 
\beqa
&&{\tl s_{13}}(z)={\rm exp}\{-\frac{i}{\sqrt k} \ph _1(z) \}, ~~
{\tl s_{32}}(z)={\rm exp}\{\frac{i}{\sqrt k} \ph _2(z) \}, \n
&&{\tl s_{24}}(z)={\rm exp}\{-\frac{i}{\sqrt k} \ph _3(z) \}.
\eeqa

\no The OPEs with the currents are  
\beqa
&&J_{31} (z) s_{13} (w)=\p _w \left(\frac{k}{z-w}
{\tl s_{13}} (w)\right),\n 
&&J_{23} (z) s_{13} (w)=\ldots, ~~~~
J_{42} (z) s_{13} (w)=\ldots, \n
&&J_{21} (z) s_{13} (w)=\p _w \left(\frac{k}{z-w}
\ps _{32}(w){\tl s_{13}} (w)\right),~~ ~~
J_{43} (z) s_{13} (w)=\ldots, \n
&&J_{41} (z) s_{13} (w)=\p _w \left(\frac{k}{z-w}
(\gm _{34}(w)-\frac{1}{2}\ps _{32}(w)\ps _{24}(w))
{\tl s_{13}} (w)\right),\n
&&J_{23} (z) s_{32} (w)=-\p _w \left(\frac{k}{z-w}
{\tl s_{32}} (w)\right),\n 
&&J_{31} (z) s_{32} (w)=\ldots, ~~~~
J_{42} (z) s_{32} (w)=\ldots, \n
&&J_{21} (z) s_{32} (w)=\p _w \left(\frac{k}{z-w}
\ps _{13}(w){\tl s_{32}} (w)\right), \n
&&J_{43} (z) s_{32} (w)=-\p _w \left(\frac{k}{z-w}
\ps _{24}(w){\tl s_{32}} (w)\right),\n
&&J_{41} (z) s_{32} (w)=-\p _w \left(\frac{k}{z-w}
\ps _{13}(w)\ps _{24}(w){\tl s_{32}} (w)\right),\n
&&J_{42} (z) s_{24} (w)=\p _w \left(\frac{k}{z-w}
{\tl s_{24}} (w)\right),\n 
&&J_{31} (z) s_{24} (w)=\ldots, ~~~~
J_{23} (z) s_{24} (w)=\ldots, \n
&&J_{43} (z) s_{24} (w)=-\p _w \left(\frac{k}{z-w}
\ps _{32}(w){\tl s_{24}} (w)\right), ~~~~
J_{12} (z) s_{24} (w)=\ldots, ~~\n
&&J_{41} (z) s_{24} (w)=-\p _w \left(\frac{k}{z-w}
(\gm _{12}(w)+\frac{1}{2}\ps _{13}(w)\ps _{32}(w))
{\tl s_{24}} (w)\right).
\eeqa
All other OPEs are trivial.

\section{Discussions}

We have studied the non-semisimple current superalgebra
$gl(2|2)^{(1)}_k$ at the general level $k$. We have constructed
its Wakimoto free field representation and Sugawara
energy-momentum tensor in the non-standard basis. We have 
found three screen currents of the first kind.

The motivation for this study is the application of the $gl(2|2)$ current
superalgebra to disordered systems. To fully take the advantage
of the CFT method, we need to construct its primary fields. 
There are two types of representations for this current superalgebra:
typical and atypical representations.
Atypical representations have no counterpart in
the bosonic algebra setting and our understanding to such representations
is still very much incomplete. So it is a highly non-trivial task
to construct primary fields corresponding to the atypical representations.
On the other hand, typical representations are similar to
those appearing in a bosonic algebra, and so  
primary fields associated with the typical representations can be
constructed by the usual procedure. Results for the primary fields will
be published in a separate paper.

\vskip 1cm

\no {\bf Acknowledgments:}

This work is financially supported by Australian Research Council. 
One of the authors (Ding) is also supported partly by the 
Natural Science Foundations of China and a Fund from AMSS.

\bebb{99}

\bibitem{BPZ}%{1}
A.A. Belavin. A. M. Polyakov, A. B. Zamolodchikov, 
{\em Nucl. Phys.} {\bf B241}, (1984)333.

\bibitem{Kac}%{2}
V. G. Kac, {\it Infinite Dimensional Lie Algebras}, third ed.,
Cambridge University press, Cambridge 1990.

\bbit{Ph}%{3}
Ph.Di Francesco, P. Mathieu and D. Senechal, {\it Conformal Field Theory}, 
Springer, 1997.

\bbit{Pol}
J. Polchinski, {\it String Theory}, Cambridge University press, 
Cambridge 1998.

\bbit{ZamA}
A. B. Zamolodchikov, {\it JETP Letters}, {\bf 43}, (1986) 730. 

%\bbit{Dot83}
%V. Dotsenko and Vl. Dotsenko, {\it Adv. Phys.} {\bf 32}, (1983) 129.

%\bbit{Lud90}
%A. Ludwig, \npb {330} {1990} 639.

%\bbit{Sha87}
%R. Shankar, \prl {58} {1987} 2466.

%\bbit{Car82}
%J.L. Cardy and S. Ostlund, {\it Phys. Rev.} {\bf B25}, (1982) 6899.

%\bbit{Lud94}
%A. Ludwig, M. Fisher, R. Shankar and G. Grinstein, {\it Phys. Rev.} {\bf
%B50}, (1994) 7526.

%\bbit{Zir94}
%M. Zirnbauer, {\it Ann. Physik} {\bf 3}, (1994) 513.

%\bbit{Cau96}
%J.-S. Caux, I.I. Kogan and A.M. Tsvelik, \npb {466} {1996} 444.

\bbit{Flohr}
M. Flohr, {\it Bits and Pieces in Logarithmic Conformal Field Theory}, 
hep-th/0111228.

\bbit{IR}
J. M. Isidro and A. V. Ramallo, \npb {414}{1994}{715}. 

\bbit{Efe83}
K. Efetov, {\it Adv. Phys.} {\bf 32}, (1983) 53. 

\bbit{Mud96}
C. Mudry, C. Chamon and X.-G. Wen, \npb {466} {1996} 383.

\bbit{BCKT}
M. J. Bhaseen,  J.-S. Caux ,  I. I. Kogan  and  A. M. Tsvelik, 
\npb {618}{2001}{465}; cond-mat/0012240.

\bbit{Ber95}
D. Bernard, hep-th/9509137.

\bbit{Maa97}
Z. Maassarani and D. Serban, \npb {489} {1997} {603}.

\bbit{Gad91}
R. Gade and F. Wegner, \npb {360} {1991} {213}.

\bbit{Gur99}
S. Guruswamy, A. LeClair and A.W.W. Ludwig, \npb {583} {2000} {475}.

\bbit{MRF}
C. Mudry, S. Ryu, A. Furusaki, cond-mat/0207723. 

%\bbit{Bha}
%M. J. Bhaseen, \npb {604}{2001}{537}; cond-mat/0011229.

\bibitem{Wak}%{}
M. Wakimoto, {\it Commun. Math. Phys. {\bf 104}}, (1986)605.

\bibitem{BOo}
M. Bershsdsky and H. Ooguri, \plb {229} {1989} {374}.
    % {\it Commun. Math. Phys.} {\bf126},(1989)49.

\bbit{FF}
 B. Feigin and E. Frenkel, \cmp {128}{1990}{161}. 

\bbit{BMP}
P. Bouwknegt, J. McCarthy and K. Pilch, {\it Prog. Phys. Suppl.} 
{\bf 102}, (1990) 67.

\bbit{Ito}
K. Ito, \plb {252}{1990}{69}

\bbit{Fren}
E. Frenkel,  hep-th/9408109.

\bbit{Bow96}
P. Bowcock, R-L.K. Koktava and A. Taormina, \plb {388} {1996} {303}.

\bbit{Ras98}
J. Rasmussen, \npb {510} {1998} {688}.

\bbit{Din01}
X.M. Ding, M.D. Gould and Y.Z. Zhang, \plb {523} {2001} {367}.

\bbit{DGMZ}
X.M. Ding, M. D. Gould, C. J. Mewton and Y. Z. Zhang, 
 hep-th/0211235.

\bbit{RS}
L. Rozanski and H. Saleur, \npb {376}{1992}{461}.

\bbit{Ras00}
J. Rasmussen, hep-th/0003035.
\eebb

\end{document}